\documentclass[amsmath, amssymb,superscriptaddress,nofootinbib]{aa}

\usepackage{graphicx}
\usepackage{amsmath}
\usepackage{amssymb}
\usepackage{xspace}
\usepackage{commath}
\usepackage{times}
\usepackage{bm} 
\usepackage{balance}
\usepackage{hyperref}
\usepackage{mathtools}
\usepackage{stmaryrd}
\usepackage{tabulary}
\usepackage{xcolor}
\usepackage{graphicx}
\usepackage{txfonts}
\usepackage{orcidlink}

\usepackage{afterpage}
\usepackage{multicol}
\newsavebox{\shortpagebox}

\makeatletter
\newcommand{\shortpage}[1]
{\par
\setbox\shortpagebox=\vbox{\strut #1\par}%
\afterpage{\onecolumn
\begin{multicols}{2}
\unvbox\AP@partial
\end{multicols}}%
\unvbox\shortpagebox
\par}
\makeatother

\hypersetup{dvips, colorlinks=true, linkcolor=blue, citecolor=blue, filecolor=blue, urlcolor=blue}

\usepackage{acronym}

\usepackage{soul} 
\usepackage{xcolor}
\definecolor{aquamarine}{rgb}{0.5, 1.0, 0.83}

\usepackage{orcidlink}
\usepackage{amsfonts}
\usepackage{color}
\usepackage{enumerate} 
\usepackage{xcolor}
\usepackage{multirow}
\usepackage{dcolumn}
\usepackage[export]{adjustbox}
\usepackage{tikz}
\usepackage{hyperref}
\usepackage[normalem]{ulem}
\usepackage{soul}
\usepackage{aasmacros}
\usepackage{amsmath}

\definecolor{aquamarine}{rgb}{0.5, 1.0, 0.83}

\begin{document}

\title{Which filaments matter: the relative scalings of anisotropic infall}

\author{Junsup Shim\inst{1,2} \orcidlink{0000-0001-7352-6175}\thanks{jsshim@pusan.ac.kr} 
\and
Dmitri Pogosyan\inst{3} \orcidlink{0000-0002-7998-6823}
\thanks{pogosyan@ualberta.ca}
\and
Myoungwon Jeon\inst{4} \orcidlink{0000-0001-6529-9777}
 \thanks{myjeon@khu.ac.kr}
\and
Christophe Pichon\inst{4,5} \orcidlink{0000-0003-0695-6735}
 \thanks{Corresponding author: pichon@iap.fr}
 }
\institute{
Department of Earth Sciences, Pusan National University, Busan 46241, Republic of Korea
\and 
Academia Sinica Institute of Astronomy and Astrophysics (ASIAA), No. 1, Section 4, Roosevelt Road, Taipei 106216, Taiwan
\and
Department of Physics, University of Alberta, 11322-89 Avenue, Edmonton, Alberta, T6G 2G7, Canada.
\and 
Department of Astronomy \& Space Science, Kyung Hee University, 1732 Deogyeong-daero, Yongin-si, Gyeonggi-do 17104, Republic of Korea
\and
Institut d'Astrophysique de Paris, 98 bis Boulevard Arago, F-75014 Paris, France
}
\date{\today}

\abstract{
Dark-matter haloes do not form in isolation but within the surrounding cosmic web. By the time a halo begins to collapse, its larger-scale environment has typically collapsed along two axes, forming filaments that channel anisotropic infall toward the halo. In this work, we derive from first principles the characteristic Lagrangian scale ratio at which such an anisotropic tidal field most strongly influences halo formation. Specifically, we identify the inflection point of the conditional probability that the tidal field, smoothed on a scale 
$R_{\rm sd}$, undergoes two-dimensional compression, given the presence of a density peak of rarity $\nu$
 on a smaller scale $R_{\rm pk}$. For a standard 
$\Lambda$CDM cosmology, we find 
$(R_\mathrm{sd}/R_\mathrm{pk})_\mathrm{infl}\approx 2.2 + (\nu-2.5)$ for $R_\mathrm{pk}$  
  corresponding to a tophat filter of 
$8{\rm Mpc}/h$. This result implies that the anisotropic tidal influence on a collapsing halo typically extends to 2–3 times the size of its Lagrangian patch. Recast as a function of formation redshift $z$, the characteristic filament scale around 
2.5$\sigma$ peaks can be approximated by
$R_\mathrm{sd}(z) \approx  31 /(2+(1+z)^2)h^{-1} {\rm Mpc}$.
We provide  practical scaling laws for selecting dynamically relevant smoothing scales in large-scale surveys and for setting initial patch sizes in high-resolution zoom simulations.
}


\maketitle
\flushbottom

\section{Introduction}\label{sec:introduction}

Large-scale cosmic filaments are massive, thread-like structures of galaxies and dark matter \citep{delapparent+86,klypin&shandarin93} that emerge as manifestations of the anisotropies of the underlying tidal field \citep{zeldovich70, bond+96}.
By channeling the anisotropic infall of matter and gas, these filaments serve as the supply lines for halos and galaxies at all redshifts, influencing their assembly histories \citep{hahn+09,borzyszkowski+17,tojeiro+17,musso+18}. Consequently, different tidal environmental factors -- including proximity to filaments \citep{musso+18}, their spatial orientation \citep{codis+15}, and local connectivity \citep{codis+18} -- drive distinct variations in the physical and dynamical properties of halos and galaxies, ranging from mass and morphology to abundance, color, luminosity, spin, and star formation \citep{aragon-calvo07, hahn+07, Sousbie2008,codis+12, libeskind+12,laigle+15, shim+15, kraljic+18, kraljic+19, laigle+18, shim&lee18, kraljic+20, song+21, lee+21, jhee+22, arora+25, yu+25}.

While tidal characteristics are recognized as key factors in shaping galaxy properties, the characteristic scale of the surrounding filamentary structure is often treated as a fixed parameter. Although multiscale filament-finding algorithms exist (e.g., DisPerSE~\citep{sousbie11}), the qualitative and quantitative variations between filaments identified at different scales relative to a given halo have not been systematically explored. In particular, the typical relative scale on which filaments contribute to halo dynamics remains poorly constrained. This leads to a fundamental question: does the influence of a filament on halos depend on its relative scale, and if so, is there a characteristic scale at which the filamentary environment most strongly governs the dynamics of haloes of a given rarity?

We address this question from first principles using Gaussian Random Field (GRF) theory, which effectively captures the statistics of cosmic fields, e.g., the underlying density and tidal fields, while relying on excursion set theory~\citep{bond+91}.
By computing the joint probability distribution functions (JPDFs) of GRF variables under specific topological constraints -- such as density peaks characterized by negative curvature -- this formalism enables a rigorous statistical description of such points.
In this paper, we draw inspiration from the statistics of critical points, i.e., points with vanishing gradient \citep{milnor63,arnold06}, extending the standard peak constraints \citep{bardeen+86} to various topological features in cosmic fields. 
This approach has not only established cosmological probes \citep{gay+12, moon+23, shim+24}, but also provided robust theoretical descriptions for the alignment \citep{codis+15}, connectivity \citep{codis+18, kraljic+22}, clustering \citep{shim+21}, and merger history \citep{cadiou+20} of the large-scale structures. 

Building on this framework, we model halos as high density peaks and filamentary tidal environments as regions that experience contraction along two axes. 
We evaluate the conditional PDF of density peaks occurring at a given location under two-dimensional tidal compression. 
This allows us to identify the preferred relative scale of filaments -- the scale ratio between filaments and halos, where filaments most sensitively influence halos. We determine this by locating the inflection point of the conditional PDF, which represents the relative scale of maximum sensitivity where the correlation changes most rapidly.

Our approach offers insights complementary to those presented in \cite{shim+21}, which investigated the clustering of critical points, including peak-to-filament statistics, and derived the characteristic length of filaments connecting clusters. While their analysis examined density peaks and saddles on a single scale at different positions, our analysis considers  high-enough density points\footnote{We employ high-density regions as statistical proxies for density peaks \citep{paranjape&sheth12} where halos form \citep{bardeen+86}. This approach allows for a more tractable analytical treatment of the conditional PDF, while maintaining a physically consistent mapping to the locations of halo formation.} and potential saddles at the {\it same} position, but  on two {\it different} scales. By focusing on this local multiscale coupling rather than spatial clustering, our framework allows us to simply identify the characteristic scale ratio at which halos (peaks) most sensitively respond to the surrounding filamentary environment (filament-saddles). 

The corresponding scaling laws should prove numerically useful for identifying the size of Lagrangian patches that need to be modeled at high resolution in zoom simulations~\citep{cadiou+22}, 
while ensuring that the anisotropy of the surrounding environment is properly taken into account. They will also help guide the analysis of observational surveys  
\citep{Euclid, LSST, DESI}, whose mass completeness should be evaluated in relation to the dynamically relevant filamentary scales.

This paper is organized as follows. Section~\ref{sec:theory} presents the theoretical description of the conditional PDF for 
 two-dimensional tidal compression  subject to density peaks occurring on smaller scales. 
Section~\ref{sec:results} presents the main results for a $\Lambda$CDM power spectrum, and discusses their local relationship  to scale-invariant power spectra,  while  Section~\ref{sec:conclusion} provides conclusions.
Details of the derivation  for  the conditional PDF for two-  and three-dimensional  GRFs are given in Appendix~\ref{appen1}.
Appendix~\ref{appen2} presents simplifications in the estimation when  power-law power spectra are assumed.
The corresponding results in two and three dimensions are given in Appendix~\ref{appen3}, while comparison to $\Lambda$CDM 
is carried in Appendix~\ref{app:lcdm}.

\section{ Predictions: halos in filaments}\label{sec:theory}

We consider the probability of having a halo at the Lagrangian scale $R_\mathrm{pk}$, determined by the halo mass, in the filamentary environment at scale $R_\mathrm{sd}$.
We propose a simple
model where the presence of the halo is determined by the overdensity value $\delta(R_\mathrm{pk})$ exceeding the critical value $\delta_c$, relying on the observation \citep{kaiser84} that high overdensity regions
track neighborhoods of density peak. Conversely,  the filamentary nature of the region is determined by the signature of the sorted eigenvalues $\lambda_3 < \lambda_2 \le 0,\lambda_1 \ge 0$ of the
deformation tensor $\Psi_{,ij}(R_\mathrm{sd})$, \textit{centred at the same position as the halo}, but at the scale $R_\mathrm{sd}$\footnote{Throughout this paper, we will use subscripts `$\mathrm{pk}$' and `$\mathrm{sd}$' as a proxy to label high density (`peak') and the signature constraint (`saddle') respectively.}. Here, $\Psi$ is the linear displacement potential of the growing adiabatic mode
\citep{zeldovich70}, scaled to satisfy $\Delta\Psi=-\delta$, while $i,j=1,2,3$ designate
the spatial derivatives\footnote{In the linear regime the displacement potential $\Psi$ is proportional to the gravitational potential $\phi$ with a
negative sign.}. 

Our primary object of interest is, thus, the conditional probability that the matter flow at scale $R_\mathrm{sd}$ is compressing along the two axes (as in a filamentary environment), given that the local overdensity $\delta$ at a scale $R_{\rm pk}$ exceeds the critical threshold $\delta_{\mathrm{pk}}$ and represents a collapsed halo. It is given by
\begin{equation}
P({\mathrm{sd}}|{\mathrm{pk}}) = \frac{P(\rm sd, {\rm pk})}{P({\rm pk})}\,, \label{eq:condPDF}
\end{equation}
where 
\begin{equation}
{P({\rm sd, pk}) \equiv
 P(\lambda_2 \le 0,\lambda_1 \ge 0 ,  \delta \ge \delta_\mathrm{pk} )
}
\end{equation}
is the joint probability of the deformation tensor and the density field subject to the relevant eigenvalues constraints, 
and
$P({\rm pk})\equiv
P(\delta \ge \delta_\mathrm{pk} )\,
$ is its marginal.

To evaluate the conditional probability in Equation~\eqref{eq:condPDF}, it is convenient to express the JPDF   using the rotational invariants introduced in~\citep{gay+12} (see Appendix~\ref{appen1}) that are scaled by their variances.
The variance of the density field smoothed on a scale $R$ is given by
\begin{equation}
\sigma_0^2(R) = 4 \pi \int dk k^2 P_m(k) W^2(k R), \label{eq:defsigma0}
\end{equation}
where $P_m(k)$ is the initial power spectrum and $W(kR)={\rm exp}(-k^2R^{2}/2)$ is the Gaussian filtering function in $k$-space with smoothing scale $R$. 
The deformation tensor is of the same order of derivatives as the density, and the variance of its trace is also $\langle \Delta \Psi(R_\mathrm{sd}) ^2 \rangle = \sigma_0^2(R_\mathrm{sd})$.

Within this formalism, the probability for the local overdensity $\delta$ at scale $R_{\rm pk}$ exceeding a critical threshold $\delta_{\rm pk}$, jointly with the condition that the matter flow at scale $R_{\rm sd}$ is compressive along two axes, can be written as follows
\begin{align} \label{eq:joinPDF}
 P({\mathrm{sd,{\rm pk}}}) &\!\!= 
\frac{25 \sqrt{5}}{12 \pi\sqrt{2\pi(1-\gamma^2)}}
 \int_{\nu_{\rm pk}}^{\infty} dy  \exp \left[-\frac{1}{2} y^2 \right]\times \\
&
\hskip -1.5cm \int_0^\infty  \!\!\!\! {\rm d} J_2
\exp\left[- \frac{5}{2} J_2 \right] 
\int_{-2\sqrt{J_2}}^{\sqrt{J_2}}  \!\!\!\! {\rm d} J_1
\left(J_1^3 - 3 J_1 J_2 + 2 J_2^{3/2} \right)
\exp\left[-\frac{1}{2} \frac{(J_1\!+\!\gamma y)^{2}}{1-\gamma^2} \right],\notag 
\end{align} 
and  its marginal
\begin{equation}
P({\rm pk})=\frac{1}{\sqrt{2\pi}}\int_{\nu_{\rm pk}}^{\infty}{\rm exp}\Big(-\frac{y^{2}}{2}\Big){\rm d}y\,. \label{eq:marginal}
\end{equation}
where $\nu_{\rm pk} \equiv \delta_\mathrm{pk}/\sigma_0({R_\mathrm{pk}})$ is a measure of rarity of a peak with overdensity $\delta_\mathrm{pk}$ at the
scale $R_\mathrm{pk}$.
For a Gaussian density field, 
a threshold $\nu_\mathrm{pk} > 2$ typically corresponds to regions around isolated maxima \citep{pogosyan+98}, whereas peaks exceeding $\nu_\mathrm{pk} > 3$ are extremely rare.\footnote{Within a spherical collapse model and standard $\Lambda$CDM cosmology, the halo of $M\approx2 \times 10^{14} h^{-1}M_\odot$ that collapses at redshift $z=0$
corresponds to $\nu_\mathrm{pk} \approx 2.1$.}

 Appendix~\ref{appen1} provides a rapid derivation of Equation~\eqref{eq:joinPDF}, including 
 details about the origin of  the boundary of integrations.
In Equation~\eqref{eq:joinPDF},   $\gamma$ is the correlation coefficient between densities at the same point in space but smoothed at scales $R_\mathrm{pk}$ and $R_\mathrm{sd}$
\begin{equation}
\gamma(R_\mathrm{pk},R_\mathrm{sd}) \!=\! \frac{\displaystyle 4 \pi\!\! \int\! dk k^2 P_m(k) W(k R_\mathrm{pk}) W( k R_\mathrm{sd})}{\sigma_0(R_\mathrm{pk}) \sigma_0(R_\mathrm{sd})}\,. \label{eq:defgamma}
\end{equation}
Together with $\nu_\mathrm{pk}$, which represents the rarity of the peak, $\gamma$ fully encodes the two physical scales under consideration, and their connections to 
the properties of the underlying primordial power spectrum.
For scale-invariant power spectra, which are a good local proxy to the $\Lambda$CDM spectrum when smoothed over a given scale, it becomes a function of the ratio between the two scales, given by Equation~\eqref{eq:defgammapowerlaw} in Appendix~\ref{appen2}.

In the next section, we will compute the realm of influence of filament  around a given peak as the scale ratio at which the conditional PDF, Equation~\eqref{eq:condPDF}, presents an inflection point as a function of $R_\mathrm{sd}$.
This is the scale that marks the transition between the volume collapsing along all three axes near the peak to the realm of two-axes collapsing flow.\footnote{We could have considered a related  definition for this realm of influence.
Indeed, nulling the second derivative of Equation~\eqref{eq:condPDF} with respect to $\gamma$ will yield some $\gamma_\mathrm{crit}(\nu_\mathrm{pk})$.
 Equating this to the $\gamma(R_\mathrm{pk},R_\mathrm{sad})$ given by Equation~\eqref{eq:defgamma} yields an implicit equation for the ratio  as a function of $\nu_\mathrm{pk}$.
For two-dimensional fields with power-law spectra, all the calculations can be done explicitly, see Appendix~\ref{appen2}. 
}
Note that a more intricate calculation would involve additionally imposing a maximum  constraint at the peak scale. This would require higher order derivatives PDFs.

\section{Results}\label{sec:results}

\subsection{Inflection versus rarity}
We study the conditional probability, $P({\rm sd|{\rm pk}})$, computed using Equation~\eqref{eq:condPDF}, as a function of the scale ratio, $R_{\rm sd}/R_{\rm pk}$. This conditional probability quantifies the likelihood that the displacement field smoothed on $R_{\rm sd}$ undergoes two-dimensional compression -- forming a filament -- given the presence of a density peak  -- a halo -- on a smaller scale $R_{\rm pk}$. As a starting point, we treat $R_\mathrm{pk}$ and $\nu_\mathrm{pk}$ as independent parameters, without
imposing the condition that the halo has collapsed.

\begin{figure}
\includegraphics[clip,width=\columnwidth]{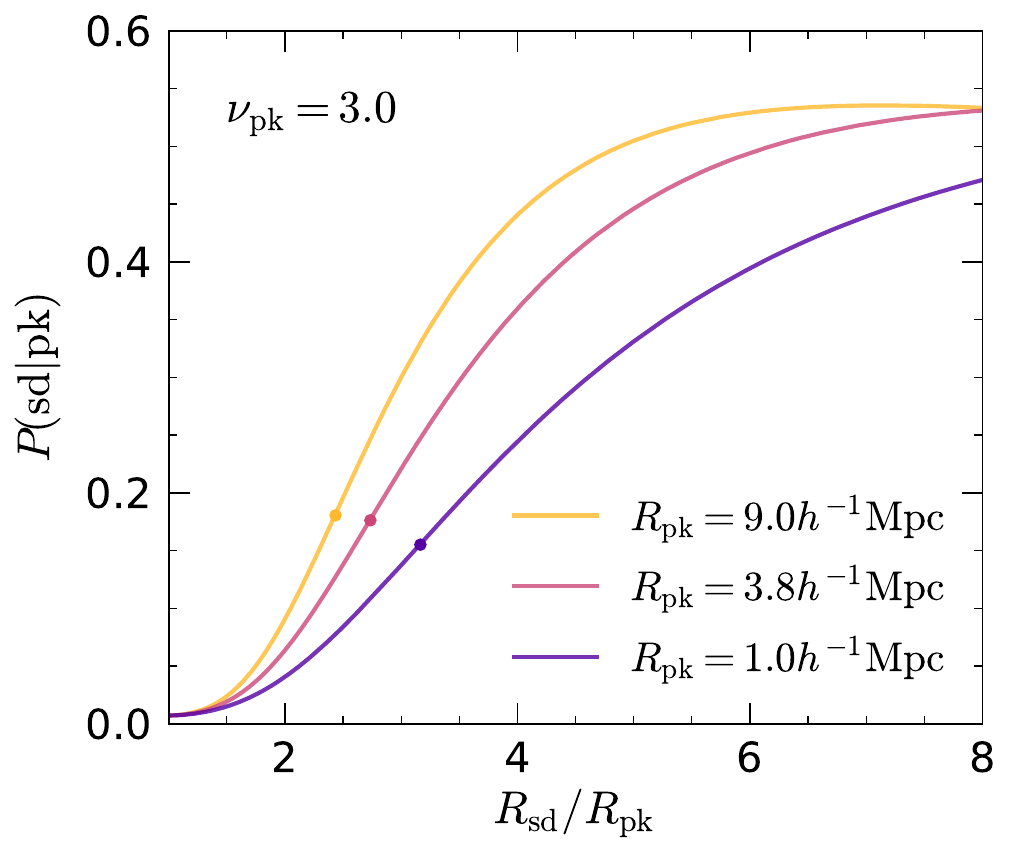}
\caption{Conditional probability (from Equation~\ref{eq:condPDF}) of identifying a filament-saddle tidal structure at scale $R_{\rm sd}$, given a density peak with ${\nu\ge\nu_{\rm pk}=3}$
at scale $R_{\rm pk}$, for a three-dimensional GRF with a $\Lambda$CDM power spectrum. Solid curves are numerical (Monte-Carlo) results, and filled circles mark the inflection points, tracing the characteristic scale ratio where the halo-filament correlation enhances most significantly. The pink curve corresponds to the tophat filter of $8h^{-1}{\rm Mpc}$, which yields $\sigma_8\approx0.81$ at $z=0$.}\label{fig:1}
\end{figure}

As a practical example, in Figure~\ref{fig:1} we consider density peaks above a fixed rarity, $\nu_\mathrm{pk}$, in a three-dimensional GRF with a $\Lambda$CDM power spectrum, smoothed on three different scales. 
Note that a Gaussian smoothing with a Lagrangian radius of $R_\mathrm{pk}=3.8h^{-1}{\rm Mpc}$ produces a density field variance equivalent to that obtained using a tophat smoothing of $8h^{-1}{\rm Mpc}$.
Thus, we relate the two scale as $R_\mathrm{pk} \approx 0.475 R_{\mathrm{pk},\textsc{th}}$.
The $P({\rm sd|{\rm pk}})$ curves are shown for 
three values of Lagrangian $R_\mathrm{pk}=1.0,3.8,$ and  $9.0 h^{-1} \mathrm{Mpc} $, which in $\Lambda$CDM model correspond to halo masses of $M=3.4\times10^{12}, 1.8\times10^{14}$ and $2.4\times10^{15}h^{-1}M_{\odot}$.

The probability $P(\mathrm{sd}|\mathrm{pk})$ is suppressed for ${R_\mathrm{sd} \sim R_\mathrm{pk}}$, since the vicinity of the peak undergoes collapse along all three directions.
It grows to the unconstrained value  $({2 \sqrt{5}+3 \tan ^{-1}\sqrt{5}})/{6 \pi } \approx 0.42 $, as $R_\mathrm{sd} \gg R_\mathrm{pk}$. This value  corresponds to the overall fraction of the volume where the filamentary condition is satisfied\footnote{For some spectra
$P({\rm sd|{\rm pk}})$ may exceed $0.42$ at intermediate values of $R_\mathrm{sd}/R_\mathrm{pk}$.}. 
We define the inflection point-- corresponding to the scale at which the growth rate of $P({\rm sd|pk})$ attains its maximum -- as the characteristic scale marking the onset of the filamentary character of the collapse around the peak. At this scale, the exclusion effect associated with the halo  constraint is largely alleviated  \citep{shim+21}, whereas  the typical  overdensity expected within a filament remains  significant.

We find that at the inflection point ${(R_{\rm sd}/R_{\rm pk})_{\rm infl}\approx2-3}$, weakly depending on the peak size $R_\mathrm{pk}$. In contrast to scale‑invariant power‑law spectra cases (see Appendix~\ref{appen2}), the slope of the power spectrum
in the $\Lambda$CDM framework steepens toward  smaller scales, which leads to larger $(R_{\rm sd}/R_{\rm pk})_{\rm infl}$ for smaller halos.

\begin{figure}
\includegraphics[clip,width=\columnwidth]{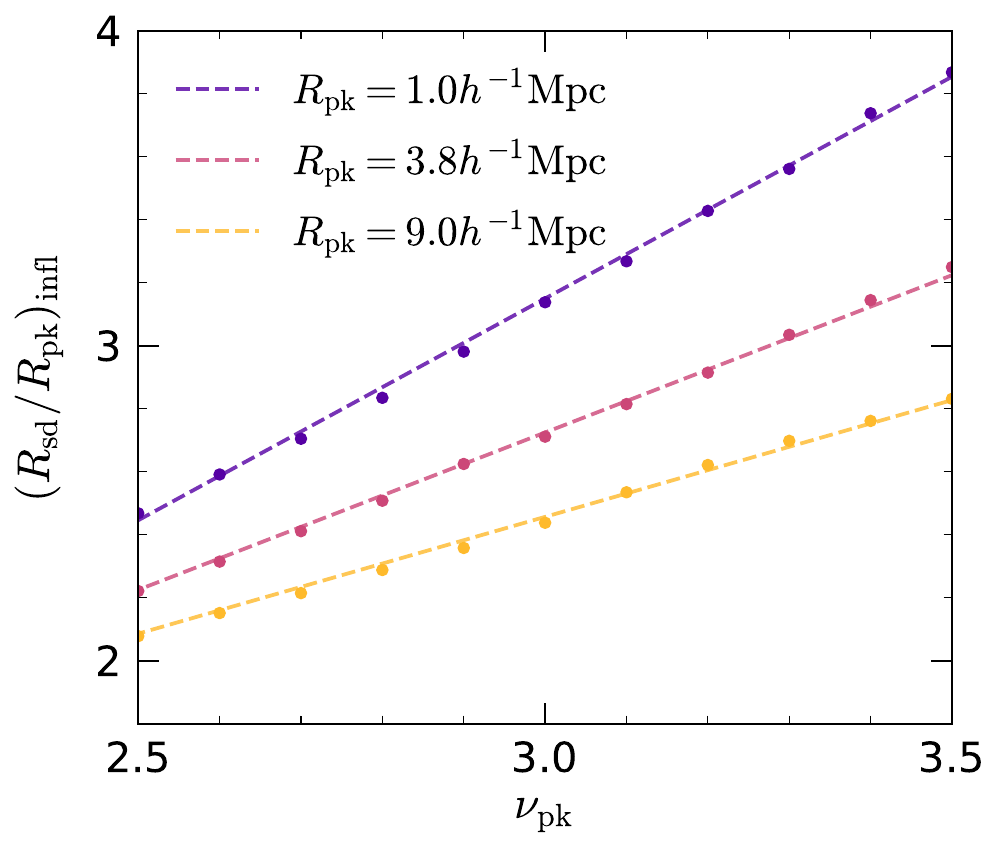}
\caption{Characteristic scale ratio as a function of peak rarity for a $\Lambda$CDM power spectrum. Dots represent numerical results at the inflection points of conditional PDFs (e.g., Figure~\ref{fig:1} for $\nu_{\rm pk}=3$), while dashed lines show linear fits. For a rarer peak, halo-filament correlation becomes most sensitive at a larger characteristic scale ratio.
}
\label{fig:2}
\end{figure}

We now show in Figure~\ref{fig:2} how the characteristic scale of filaments, i.e., $(R_{\rm sd}/R_{\rm pk})_{\rm infl}$, depends on peak rarity. We find that it increases monotonically with peak rarity. 
Typically, we find ${2 \le (R_{\rm sd}/R_{\rm pk})_{\rm infl} < 4}$, with higher peaks exhibiting a larger zone of influence and consequently a greater value of $(R_{\rm sd}/R_{\rm pk})_{\rm infl}$.
Conversely, for a fixed peak rarity, the trend of increasing $(R_{\rm sd}/R_{\rm pk})_{\rm infl}$ toward smaller $R_{\rm pk}$ persists across all values of $\nu_{\rm pk}$.

The relation between the peak rarity 
and the characteristic scale ratio 
at the inflection points is well described by a linear fit in $\nu_\mathrm{pk}$,
\begin{align}
\label{eqn:linear}
    (R_{\rm sd}/R_{\rm pk})_{\rm infl}\!=\!&\left(\left[\!\frac{R_{\rm pk}}{9.1 \mathrm{Mpc}/h}\right]^2 \!-\! \frac{R_{\rm pk}}{4.9 \mathrm{Mpc}/h}\!+\! 1.6 \right)\!\nu_{\rm pk}\notag\\
    & -\left(\left[\!\frac{R_{\rm pk}}{6.5 \mathrm{Mpc}/h}\right]^2 \!-\! \frac{R_{\rm pk}}{2.5\mathrm{Mpc}/h} \!+\! 1.45\right).
\end{align}

This linear parameterization reproduces the characteristic scale ratio with an accuracy better than $1\%$ of the measured values for $2.5 \leq \nu_{\rm pk} < 3.5$ across all three peak sizes examined.
As an illustrative example, for peaks identified using Gaussian smoothing with $R_{\rm pk} = 3.8,h^{-1}{\rm Mpc}$ —equivalently, an $8\,h^{-1}{\rm Mpc}$ tophat smoothing—the direct fit is well described by a practical rule of thumb
\begin{equation}
(R_{\rm sd}/R_{\rm pk})_{\rm infl}\approx 2.2 +(\nu_{\rm pk}-2.5),
\label{eq:Rsdofnu}
\end{equation}
which remains accurate within $1\%$. 
This is our first bring-home rule.

\subsection{Astrophysical applications}
We now examine the filament scales around peaks that have formed collapsed halos.
We can then rely on the spherical collapse criterion
\begin{equation}
\nu_\mathrm{pk} \sigma_0(R_\mathrm{pk})  D(z) = \delta_c~,
\label{eq:sphere_collapse}
\end{equation}
where $D(z)$ is a growing mode of cosmological perturbations, normalized
at the present time, $D(0)=1$, and $\delta_c=1.686$ is the critical linear overdensity. This condition establishes
a relationship between the Lagrangian scale, $R_\mathrm{pk}$, and the rarity of a halo, $\nu_\mathrm{pk}$, and redshift $z$.

\subsubsection{$R_\mathrm{sd}$ of collapsed objects versus mass and redshift}
Let us first focus on $R_\mathrm{sd}/R_\mathrm{pk}$ as a function of collapsing halo size at fixed redshift.
For a $\Lambda$CDM spectrum, Equations~\eqref{eq:defsigma0} and~\eqref{eq:sphere_collapse} define  the 
relationship
\begin{equation}
\nu_\mathrm{pk}(R_\mathrm{pk},z)\! \approx\! \frac{0.8}{D(z)\sigma_8} \! \left(\! \left[\frac{R_\mathrm{pk}}{5.15 \mathrm{Mpc}/h}\! \right]^{\frac{1}{4}}
\!\!\!+ \frac{R_\mathrm{pk}}{3.14 \mathrm{Mpc}/h}\!\right), 
\label{eq:nuofRpk}
\end{equation}
in the range of $R_\mathrm{pk}$ from $0.1$  to $10~h^{-1}\mathrm{ Mpc}$ that 
covers six orders of magnitude in mass from $M \approx 3.3\times 10^9$ to $M \approx 3.3 \times 10^{15} h^{-1}\mathrm{M_\odot} $. 
Conversely, the Gaussian Lagrangian scale of a halo is determined by its mass
\begin{equation}
R_\mathrm{pk} \approx 3.1 \; M_{14}^{1/3} \;
h^{-1} \mathrm{Mpc}~,
\label{eq:mass_to_radius}
\end{equation}
where ${M_{X} = M/( 10^{X} (\Omega_m/0.3) h^{-1} \mathrm{M_\odot})}$. 
Substituting 
Equation~(\ref{eq:nuofRpk}) into Equation~(\ref{eqn:linear}) gives the relative filamentary scales  $R_\mathrm{sd}/R_\mathrm{pk}$ as a function 
of $R_\mathrm{pk}$ for collapsed halos
at different redshifts.  

\begin{figure}
\includegraphics[clip,width=\columnwidth]{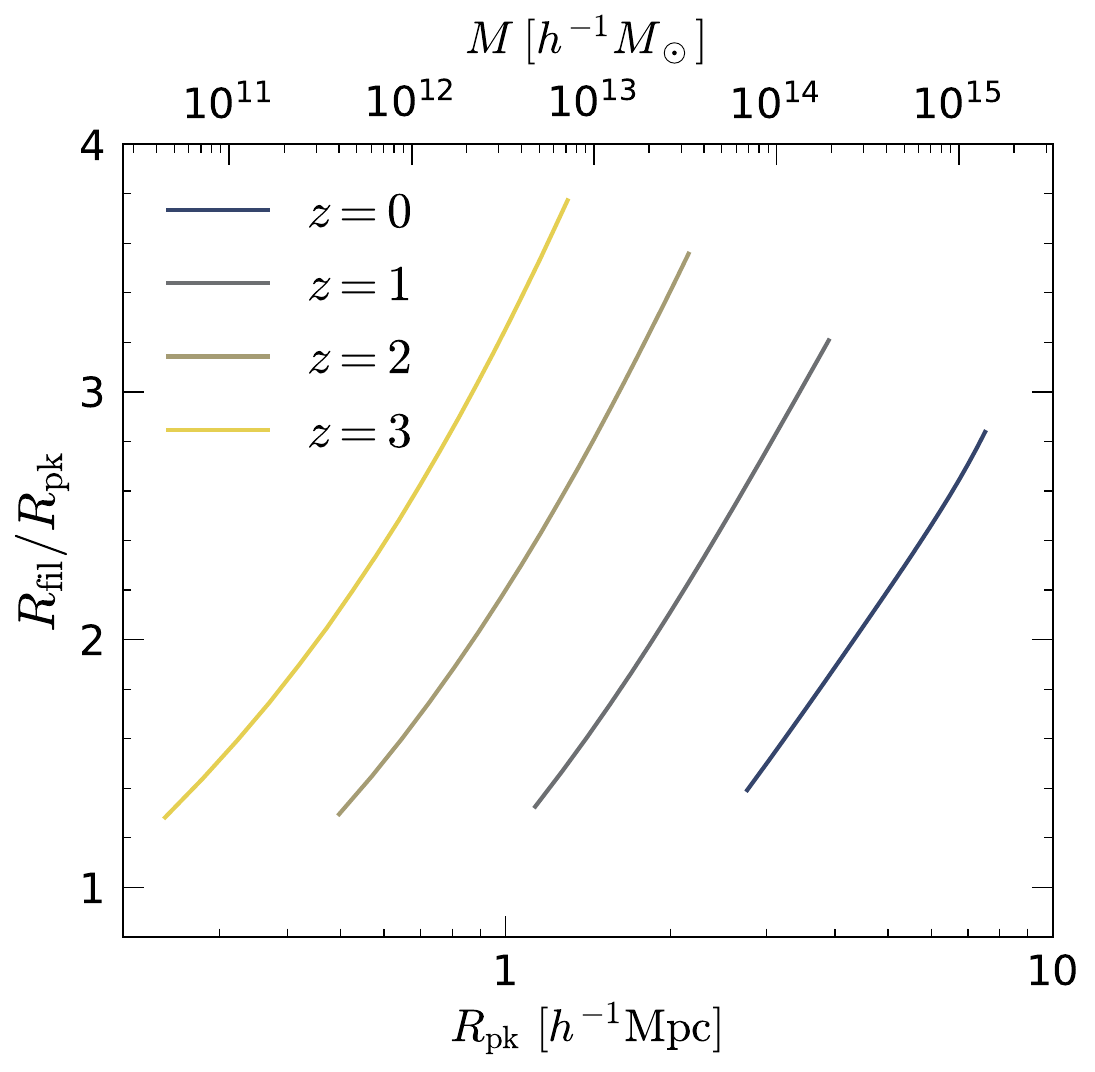}
\caption{Characteristic filamentary scale ratio around halos on various scales that collapsed at various redshifts.
Halo masses (top $x$-axis) corresponding to the peak scales (bottom $x$-axis) are also marked. 
}\label{fig:cmax_Rpk}
\end{figure}

Figure~\ref{fig:cmax_Rpk} presents these curves for $z=0,1$, and $2$ in a $\Lambda$CDM universe with $\sigma_8=0.81$.\footnote{We set the condition ${1.7 < \nu_\mathrm{pk} < 3.5}$, the range where the high-field value is still a good approximation for the peak condition,
but the peaks are not exceedingly rare. This defines the redshift dependent range of scales under consideration. 
At ${z=0}$ the covered range for halo Lagrangian size is, thus, ${R_\mathrm{pk} \approx 2.8\text{–}7.5\, h^{-1}\mathrm{Mpc}}$, corresponding to ${M \approx 7 \times 10^{13}\text{–}1.4 \times 10^{15}\, h^{-1} M_\odot}$.} We observe a significant change in the relative filament scale with increasing mass, which can be explained by the transition from abundant low-mass halos to rare high-mass ones. The tendency is even more pronounced at high redshifts where the slope of the power spectrum is steeper in the relevant range of scales.
In the range of relevant scales, the  curves in Figure~\ref{fig:cmax_Rpk} can be fitted by simple power-laws\footnote{Note that the power index in the fit corresponds to $  (n_\mathrm{eff}+3)/2\approx 0.68$ for an effective 
slope of the power spectrum, $n_\mathrm{eff} \approx -1.6$ at $3.8 h^{-1}\mathrm{Mpc}$ (see Figure~\ref{fig:neff} in Appendix~\ref{app:lcdm}). 
}
\begin{subequations}\label{eq:z0fit}
\begin{align}
(R_\mathrm{fil}/R_\mathrm{pk}) &\approx (R_\mathrm{pk}/1.6 h^{-1} \mathrm{Mpc})^{0.68}\quad \,\,\,\mathrm{for}\quad {z=0},\\
(R_\mathrm{fil}/R_\mathrm{pk}) &\approx (R_\mathrm{pk}/0.67 h^{-1} \mathrm{Mpc})^{0.68}\quad \mathrm{for}\quad {z=1},\\
(R_\mathrm{fil}/R_\mathrm{pk}) &\approx (R_\mathrm{pk}/0.32 h^{-1} \mathrm{Mpc})^{0.68}\quad \mathrm{for}\quad {z=2},\\
(R_\mathrm{fil}/R_\mathrm{pk}) &\approx (R_\mathrm{pk}/0.15 h^{-1} \mathrm{Mpc})^{0.62}\quad \mathrm{for}\quad {z=3},
\end{align}
\end{subequations}
or, using  mass in place of $R_\mathrm{pk}$ from Equation~(\ref{eq:mass_to_radius}),
\begin{subequations}\label{eq:massfil}
\begin{align}
R_\mathrm{fil}(M) &\approx 4.9  \; M_{14}^{0.56} \; h^{-1} \mathrm{Mpc}\quad \mathrm{for}\quad {z=0}\,, \\
R_\mathrm{fil}(M) &\approx 2.4  \; M_{13}^{0.56} \; h^{-1} \mathrm{Mpc}\quad \mathrm{for}\quad {z=1}\,, \\
R_\mathrm{fil}(M) &\approx 1.1  \; M_{12}^{0.56} \; h^{-1} \mathrm{Mpc}\quad \mathrm{for}\quad {z=2}\,,\\ 
R_\mathrm{fil}(M) &\approx 0.49  \; M_{11}^{0.54} \; h^{-1} \mathrm{Mpc}\quad \mathrm{for}\quad {z=3}\,. 
\end{align}
\end{subequations}
More generally, for arbitrary redshifts, substituting Equation~\eqref{eq:mass_to_radius} into Equations~\eqref{eq:nuofRpk}
and~\eqref{eqn:linear} allows us to simply compute $R_\mathrm{fil}(M,z)$.
This is our second bring-home rule.

\subsubsection{The environment of the most prominent halos}

We now consider the redshift evolution of the characteristic scale of filaments associated with the formation of the largest
halos at each redshift.  Thus, we fix the halo rarity $\nu_\mathrm{pk}$ threshold at a high value, find from the collapse condition in Equation~(\ref{eq:sphere_collapse}) the Lagrangian size of such halos as a function of the redshift, $R_\mathrm{pk}(z)$,  and, subsequently, determine the  $R_\mathrm{sd}(z)$ for these $R_\mathrm{pk}(z)$ using Equation~\eqref{eqn:linear}.  

\begin{figure}
\includegraphics[clip,width=\columnwidth]{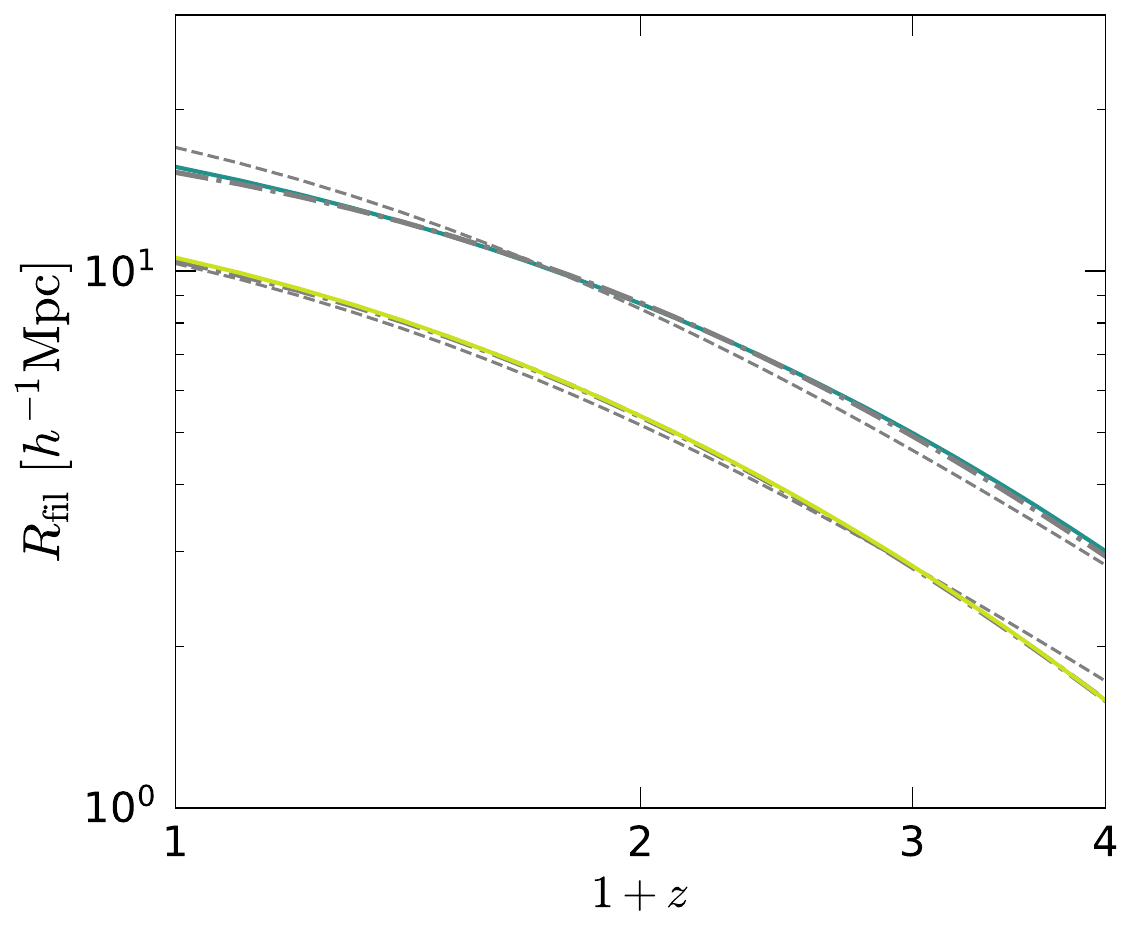}
\caption{
Characteristic scale of the filamentary tidal environment that most strongly influences density peaks at the time of their collapse, as given by Equation~\eqref{eqn:linear}. Results are shown for peaks with $\nu_{\rm pk} = 3.0$ and $2.5$ (solid lines), together with their corresponding fits (dashed lines from Equation~\eqref{eq:Rdcase} and dash-dotted lines from Equation~\eqref{eq:Rdzprecise}). This relation identifies the characteristic filament scale that governs the formation of the most massive halos at their epoch of formation.
}\label{fig:3}
\end{figure}

In Figure~\ref{fig:3}, we consider $\nu_{\rm pk}=2.5$ and $3.0$ density peaks that have reached the spherical collapse threshold density at various epochs, representing some of the most massive halos that formed at any given redshift.
For $\nu_{\rm pk}=2.5$, these rarest halos typically have masses of $M\approx3.9\times 10^{14}, 4.1\times 10^{13}, 5.3\times10^{12}$ and  $1.8\times10^{11} h^{-1}M_{\odot}$, at $z=0, 1, 2$ and  $4$ respectively.

The characteristic scale of filaments increases toward the present day, reflecting the hierarchical assembly of cosmic structures. As more massive halos form at later epochs, the filaments associated with them correspondingly grow larger. This main trend is complemented by the change of the 
effective slope of the power spectrum at the progressively increasing scale of collapsing objects. We find that the overall trend is well captured with the following simple fit,
\begin{align}
    R_{\rm fil} \,[h^{-1}{\rm Mpc}] = 
\begin{cases} 
\displaystyle \frac{31}{2+(1+z)^2} & \text{for }  \nu_{\rm pk}=2.5,\\[1.5ex]
\displaystyle \frac{51}{2+(1+z)^2} & \text{for } \nu_{\rm pk}=3.0, 
\end{cases}  \label{eq:Rdcase}
\end{align}
which remains accurate within $10\%$ at least up to $z\approx3$. 
This is our third bring-home rule.
We also provide a more sophisticated theoretically-motivated
fitting form in Appendix~\ref{appen5}, which maintains better accuracy.

\subsubsection{Mapping Lagrangian to Eulerian scales} Let us translate our findings into a procedure to optimally smooth observational surveys so as to preserve the dynamically relevant cosmic environment around a given halo. 
For a halo of a given Lagrangian scale, $R_\mathrm{pk}$ (resp. mass, $M$), we can apply the procedure described below Equation~\eqref{eq:massfil} to compute the corresponding Lagrangian scale, $R_{\rm fil}(M,z)$. Conversely, for a halo of a given rarity, we should turn to Equation~\eqref{eq:Rdcase}. 
We suggest that the  corresponding smoothing scale in  Eulerian space should   be chosen to be  $R_{\rm fil,Eulerian} \equiv R_{\rm fil,Lagrangian} $, because filaments lie in mildly nonlinear regime, where the difference between the Lagrangian and Eulerian scales is not significant.\footnote{For instance, focusing on $10^{14}\ h^{-1}{\rm M}_\odot$ halos in a vanilla $\Lambda$CDM universe at $z=0$, Equation~\eqref{eq:massfil} states that the large scales structures should be smoothed over a $R_{\rm fil,Eulerian} = 4.9 h^{-1} \mathrm{Mpc}$ Gaussian filter. }

\section{Conclusion\label{sec:conclusion}}
We investigated whether a characteristic scale exists at which filamentary environments most strongly influence the formation of halos and galaxies. To this end, we analyzed the conditional probability distribution function of the density of potential saddles -- that delineate filamentary environments, conditioned  to the density peaks --tracers of halos. We defined the characteristic scale ratio $(R_{\rm sd}/R_{\rm pk})_{\rm infl}$ between these density and tidal structures as the inflection point of the conditional PDF, marking the regime where the halo– filament correlation is maximally sensitive to variations in the filament scale.

Our results show that, although the strength of the halo–filament correlation increases monotonically with the scale ratio, its sensitivity reaches a maximum at a specific value dependent on both the rarity, $\nu_{\rm pk}$, and the scale, $R_{\rm pk}$, of the density peak. This implies that the \emph{effective} filamentary environment does not correspond to a fixed, universal scale but is closely tied to the peak’s scale and rarity—that is, to halo mass. Our findings for the three-dimensional $\Lambda$CDM power spectrum are consistent with those obtained for power-law spectra (see Appendix~\ref{appen4}) and remain robust in the two-dimensional analyses where analytical predictions are available (see Appendix~\ref{appen31}).

Our findings are therefore summarized as follows:
    the characteristic scale ratio which halo formation responds most sensitively to filaments varies with halo scale (i.e., size and rarity of a peak),
     following Equation~\eqref{eqn:linear};
     the linear scaling establishes the zone of anisotropic influence on collapsing halos, typically extending to $2$--$3$ times the size of their Lagrangian patch;
     the typical filament scale one should consider as a function of peak scale or mass and redshift is given by Equations~\eqref{eq:z0fit} and~\eqref{eq:massfil};
    the characteristic filament scale evolves with redshift and is well approximated by Equation~\eqref{eq:Rdcase}.

These findings offer practical guidance for identifying the characteristic scales of filamentary structures in galaxy evolution studies.
The derived scaling relations provide a framework for selecting optimal smoothing lengths in survey analyses on the one hand \citep{laigle_etal_2025}, and for defining the initial patch sizes in zoom-in simulations \citep{Buehlmann_2025}, ensuring that the relevant filamentary environment is accurately captured on the other hand.

Indeed, our result offers a quantitative improvement over the usual practice of choosing buffer regions in zoom simulations by trial and error or purely based on spherical overdensity arguments, and it provides a clean way to control systematics in studies of spin alignment, shape, and merger histories driven by anisotropic infall.
 
Conversely, in large surveys such as Euclid, LSST, or DESI \citep{Euclid, LSST, DESI}, one can tune multiscale filament finders so that the smoothing length used to define filaments corresponds to the dynamically relevant scale inferred for the haloes and galaxies under study, rather than adopting a one-size-fits-all scale.  In such observational  settings one usually deals with Eulerian scales.   When one deals
with  relatively rare halos, these scales are in mildly nonlinear regime, where  the Lagrangian and Eulerian
scales are similar. So our recipe is to determine the scale of the peaks from their mass, and using $R_\mathrm{sd}$ in Eulerian space.
This is especially important when interpreting environmental trends in galaxy properties (e.g. colour, morphology, or star formation) as a function of distance to filaments, since  the ``right'' filament scale depends on halo rarity and evolves with redshift. 
\\
\section*{Acknowledgements}
We thank Corentin Cadiou,  Simon Prunet, Ho Seong Hwang, Katarina Kraljic
and Sabri Errachdi for useful discussions.
JS acknowledges the support by
Academia Sinica Institute of Astronomy and Astrophysics. M.-J. is supported by Samsung Science and Technology Foundation under Project Number SSTF-BA2402-03.
This work is partially supported by the grant
GALBAR ANR-25-CE31-4684 and  IDF-DIM-ORIGINES-2023-4-11,
and from the CNRS through the MITI interdisciplinary programs.
\bibliographystyle{aa}
\bibliography{main}

\appendix

\section{Rotational invariants and PDFs}\label{appen1}
Here we present the Gaussian joint probability distribution function (JPDF)
of the (scaled) deformation tensor and the density, defined at the same point in space,
but at different scales.

\subsection{Rotational invariants in 3D}\label{appen11}

It is convenient to write the JPDF of the relevant quantities using variables normalized by the density variance,
$x_{ij}=\Psi_{ij}/\sigma_0(R_{\rm sd})$ and $y = \delta/\sigma_0(R_{\rm pk})$. 
The JPDF  then has a similar formal structure
\begin{equation}
P(x_{ij}, y) d x_{ij} dy\,,
\end{equation}
as the JPDF of the field
and its derivatives studied in \citep{gay+12}. The only difference is that in \citep{gay+12} $x_{ij}$ represented 
the normalized version of the Hessian of the density, and here it is the deformation tensor. 
Both the density Hessian and deformation tensor eigenvalues have the same filamentary condition $\lambda_3 \le \lambda_2 < 0, \lambda_1 > 0$,
and both are negatively correlated with the density, $\langle \mathrm{Tr}( x_{ij} ) \delta  \rangle = -\gamma$. Correspondigly,  in \citep{gay+12}
$\gamma$ represents the  correlation between density and its Laplacian at the same scale, whereas in this paper it represents the  density correlation at different scales but at the same point.
We are once again interested in statistical measures that are rotation-invariant at arbitrary points in space. Therefore, the results of \citep{gay+12}, including the development of non-Gaussian corrections, can be directly applied in the present work.

Following \citep{gay+12}, we introduce  polynomial in field rotational invariants that
describe our problem:   the density value $y$ itself
and the invariants of the second rank matrix $x_{ij}$.
\begin{subequations}
\begin{align}
I_1 &\equiv \mathrm{Tr}(x_{ij}) = x_{11}+x_{22}+x_{33} 
\,, \\
I_2 &\equiv x_{11}x_{22}\!+\!x_{22}x_{33}\!+\! x_{11}x_{33}
\!-\! x_{12}^2 \!-\! x_{23}^2 \!-\! x_{13}^2 
\,,\\
I_3 &\equiv \det\left| x_{ij} \right|  = 
x_{11} x_{22} x_{33} 
+ 2 x_{12} x_{23} x_{13} -  \notag\\
& \quad\quad\quad\quad\quad\quad - x_{11} x_{23}^2 - x_{22} x_{13}^2 - x_{33} x_{12}^2 \,.
\end{align}
\end{subequations}
The same invariants expressed through eigenvalues are
\begin{equation}
I_1  \!=\!
\lambda_1+\lambda_2+\lambda_3 
\,,~
I_2 \!\equiv\! 
 \lambda_1 \lambda_2 + \lambda_2 \lambda_3 + \lambda_1 \lambda_3 
\,,~
I_3 \!=\! \lambda_1 \lambda_2 \lambda_3 
\,. \notag
\end{equation}
These variables can be made partially independent using the linear combinations
\begin{equation}
J_1  =  I_1 , ~~
J_2  =  I_1^2 - 3 I_2 ~~{\rm and} ~~
J_3  =  I_1^3 - \frac{9}{2} I_1 I_2 + \frac{27}{2} I_3, \notag
\end{equation}
which with our normalization have the following lowest (up to quadratic
in the field variables) moments
\begin{equation}
 \langle y \rangle = 0, ~~ \langle J_1 \rangle = 0\,, ~~ 
\langle y^2 \rangle =1, ~~ 
\langle {J_1}^2\rangle =1,
~~ \langle J_2 \rangle = 1 \,, \notag
\label{eq:varnorm}
\end{equation}
and the only non-vanishing cross-correlation
 \begin{equation}
 \quad \langle y J_1 \rangle = -\gamma(R_{\rm pk},R_{\rm sd})\,,
\end{equation}
where
$\gamma(R_{\rm pk},R_{\rm sd})$
is the correlation coefficient between densities at the same point in space but smoothed at scales $R_{\rm pk}$ and $R_{\rm sd}$ given by Equation~\eqref{eq:defgamma}. Clearly, $\gamma(R,R)=1$.
In terms of these variables, the Gaussian JPDF  is given by
\begin{equation}
\label{eqn:10}
 G_{\rm 3D} (y, J_1, J_2, J_3) \!=\! 
\frac{25 \sqrt{5}\displaystyle \exp\left[-\frac{1}{2} \frac{y^2 \!+\! 2 \gamma y J_1 \!+\! J_1^2}{1-\gamma^2}
- \frac{5 J_2}{2}  \right]}{12 \pi\sqrt{2 \pi (1\!-\!\gamma^2)}} 
\,, \notag
\end{equation}
with $J_3$ distributed uniformly between
$-J_2^{3/2}$ and $J_2^{3/2}$.

Then, by definition, $P({\rm sd},{\rm pk}) $ reads
\begin{equation}
P({\rm sd},{\rm pk}) \equiv \label{eq:joinPDF3D}
\langle \theta(-\lambda_2 ) \theta(\lambda_2-\lambda_3 ) \theta(\lambda_1 )  \theta(\nu-\nu_\mathrm{pk}) \rangle \,,   
\end{equation}
 requiring the flow to be compressive along two axes and the peak density contrast to be above the threshold $\nu_\mathrm{pk}$.
In Equation~\eqref{eq:joinPDF3D}, the expectation is evaluated using the Gaussian JPDF $G_{\rm 3D}$,
while $\theta$ is the Heaviside function.
The filamentary domain for the deformation tensor, expressed in terms of the $J_i$ variables, \citep[see, e.g.][]{gay+12} reads
$J_2 \in [0,\infty[$, $J_1 \in [-2 J_2^{1/2},J_2^{1/2}]$ and
$J_3 \in [-\frac{1}{2} J_1^3 + \frac{3}{2} J_1 J_2, J_2^{3/2}]$. 
 Hence
\begin{equation}
P({\rm sd},{\rm pk})\!=\!  \int_{\nu_{\rm pk}}^{\infty}{\rm d}y\int_{0}^{\infty}\!\!\!\! \mathrm{d}J_{2}\int_{-2 J_2^{1/2}}^{J_2^{1/2}} \!\!\!\!\! \mathrm{d}J_{1}  \int_{-\frac{1}{2} J_1^3 + \frac{3}{2} J_1 J_2}^{ J_2^{3/2}}\!\!\!\!\!\!\!\! \mathrm{d}J_{3}  \, G_{\rm 3D}\,,\notag
\end{equation}
leading to Equation~\eqref{eq:joinPDF} in the main text after integration over $J_3$.\footnote{The integral over $J_1$ can also be performed analytically, although the resulting expression is rather cumbersome.} 

\subsection{Rotational invariants in 2D}\label{appen12}
The rotation invariant formalism is straightforwardly adapted to  2D random 
fields,  which are of interest for e.g. CMB,
intensity maps and weak lensing studies.
There are just two polynomial invariants that describe the deformation tensor in 2D,
\begin{equation}
    I_1 \equiv \mathrm{Tr}(x_{ij}) = x_{11}+x_{22}\,\,\,{\rm and}\ 
    I_2 \equiv \det\left| x_{ij} \right| = x_{11} x_{22} - x_{12}^2\, \notag
\end{equation}
expressed via eigenvalues as
\begin{equation}
I_1 \equiv \mathrm{Tr}(x_{ij}) =
\lambda_1+\lambda_2 
\,, \quad{\rm and} \quad
I_2 \equiv \det\left| x_{ij} \right| = \lambda_1 \lambda_2
\,. \notag
\end{equation}
We shall also use the orthogonal set
\begin{equation}
J_1  =  I_1 ,  \quad{\rm and} \quad
J_2  =  I_1^2 - 4 I_2 = \left( x_{11}-x_{22} \right )^2 + 4 x_{12}^2  \notag
.
\end{equation}
We can then express the 2D Gaussian JPDF including the field value, $y$,
normalized over ${\rm d}y {\rm d}J_1 {\rm d}J_2$ as 
\begin{equation}\label{eqn:2d}
G_{\rm 2D}(y,J_1, J_2)\!\!=\! \!\frac{\exp\left[\displaystyle-\frac{1}{2} \frac{y^{2}\!+\!2\gamma y J_{1}\!+\! J_{1}^{2}}{1-\gamma^2} \!-\! J_{2}
\right]}{2 \pi \sqrt{1-\gamma^2}} \!
. 
\end{equation}
The integration domain associated with the saddle condition in the velocity potential field is defined by \citep{gay+12}
as $J_1 \in [-\infty, \infty]$, $J_2 \in [J_1^2, \infty]$, with the rarity condition, $y \ge\nu_{\rm pk}$.
In contrast to 3D case, the integration in Equation~\eqref{eqn:2d} over $J_{1}$, $J_{2}$, and $y$ can be done analytically, and yields
\begin{equation}
P_{\rm 2D}(\rm sd,{\rm pk})\!\equiv\!
\langle \theta(-\lambda_2 )\theta(\lambda_1 )  \theta(\nu-\nu_\mathrm{pk}) \rangle
\!=\!\frac{1}{2\sqrt{3}} \operatorname{erfc}{\left(\frac{\sqrt{3} \nu_{\rm pk}}{ \sqrt{6 - 4 \gamma^{2}}} \right)}, \notag
\end{equation}
with the marginal
\begin{equation}
P({\rm pk})=\frac{1}{2}{\rm erfc}\Big(\frac{\nu_{\rm pk}}{\sqrt{2}}\Big),
\end{equation}
which allows us to   finally write the conditional PDF
in the final purely analytic form
\begin{equation}\label{eqn:condPDF2d}
P_{\rm 2D}(\rm sd|{pk})=\frac{P_{\rm 2D}(\rm sd,{pk})}{P({pk})}=
\frac{1}{\sqrt{3}}\frac{{\rm erfc\Big(\frac{ \sqrt{3}\nu_{\rm pk}}{\sqrt{6-4\gamma^{2}}}\Big)}}{{\rm erfc}\Big(\frac{\nu_{\rm pk}}{\sqrt{2}}\Big)}.
\end{equation}
%

\section{Power-law spectra}\label{appen2}
In this and the following two appendices, we perform  validity tests of our numerical calculations, 
using scale-invariant power-law spectra, $P_m(k)\propto k^{n}$. For such spectra 
$\sigma_0(R) \sim R^{-\frac{n+d}{2}}$ and
the correlation coefficient in Equation~\eqref{eq:defgamma} becomes a simple function of two variables -- the scale ratio $R_{\rm sd}/R_{\rm pk}$ and spectral index $n$ given by
\begin{equation}
    \gamma(R_{\rm sd},R_{\rm pk})=\bigg(\frac{2(R_{\rm sd}/R_{\rm pk})}{1+(R_{\rm sd}/R_{\rm pk})^{2}}\bigg)^{\frac{n+d}{2}},
    \label{eq:defgammapowerlaw}
\end{equation}
where $d$ is the dimension of the space and  $ n > -d$.

For the power law spectrum, $\gamma(R_\mathrm{sd},R_\mathrm{pk})$ clearly depends only on the ratio $R_\mathrm{sd}/R_\mathrm{pk}$. As the result,
any characteristic scale $R_\mathrm{sd}$ of the probability distribution $P(\mathrm{sd}|\mathrm{pk})$ (treated as an implicit function of $R_\mathrm{sd}$ via $\gamma$) corresponds to some characteristic value, $\gamma_*$
\begin{equation}
 \gamma(R_{\rm sd}/R_{\rm pk}) = \gamma_*(\nu_\mathrm{pk},n,d)
 \label{eq:gammastar}
\end{equation}
that does not depend on $R_\mathrm{pk}$.  Then, the  $\gamma_*$ corresponding to the inflection point, 
${d^2 P(\mathrm{sd}|\nu_\mathrm{pk})}/{d R_\mathrm{sd}^2}=0$, is
defined implicitly by the equation
\begin{equation}
    \gamma  \frac{d^2 P
    }{d^2 \gamma}  + 
 \frac{d P
 }{d \gamma} \!
\left(\! 1\!-\!\frac{\gamma^{\frac{4}{n+d}}-\sqrt{1\!-\!\gamma^{\frac{4}{n+d}}}}{\frac{n+d}{2} \left(1\!-\!\gamma^{\frac{4}{n+d}}\right)}\right)\!=\! 0.\label{eq:implicit}
\end{equation}
Once $\gamma_*$ is identified, inverting Equation~(\ref{eq:gammastar}) for the scale ratio gives
\begin{equation}
R_\mathrm{sd}/R_\mathrm{pk} = \gamma_*^{-\frac{2}{n+d}} \left(1+ \sqrt{1-\gamma_*^{\frac{4}{n+d}}} \right)\,. \label{eq:Rimplicit}
\end{equation}
For instance, for 2D scale invariant spectra and $\nu_\mathrm{pk}=2.5$,
we find, via Equations~\eqref{eq:implicit} and~\eqref{eq:Rimplicit}, and given the analytical Equation~(\ref{eqn:condPDF2d})  for $P(\mathrm{sd}|\mathrm{pk})$,
$\gamma_* \approx 0.79, 0.73, 0.69$ and $R_\mathrm{sd}/R_\mathrm{pk} \approx 4.8,3.5,2.9$ for $n=-1.5,-1,-0.5$ respectively. 
The $R_\mathrm{sd}$ scale raises up to $R_\mathrm{sd}/R_\mathrm{pk} \approx 7.1,4.5,3.5$ in the environment of rare peaks, $\nu_\mathrm{pk}=3$.

In Appendix~\ref{appen3} we present the 2D and 3D GRF power-law cases, and  compare the 3D case to the $\Lambda$CDM model
presented in the main text in Appendix~\ref{appen5}.

\section{Power-law spectra zone of influence}\label{appen3}
Let us present here the zone of influence of peaks in scale-invariant Gaussian random fields,
for which analytical results are available in 2D.

\subsection{2D Power-law spectra}\label{appen31}

\begin{figure}
\includegraphics[clip,width=\columnwidth]{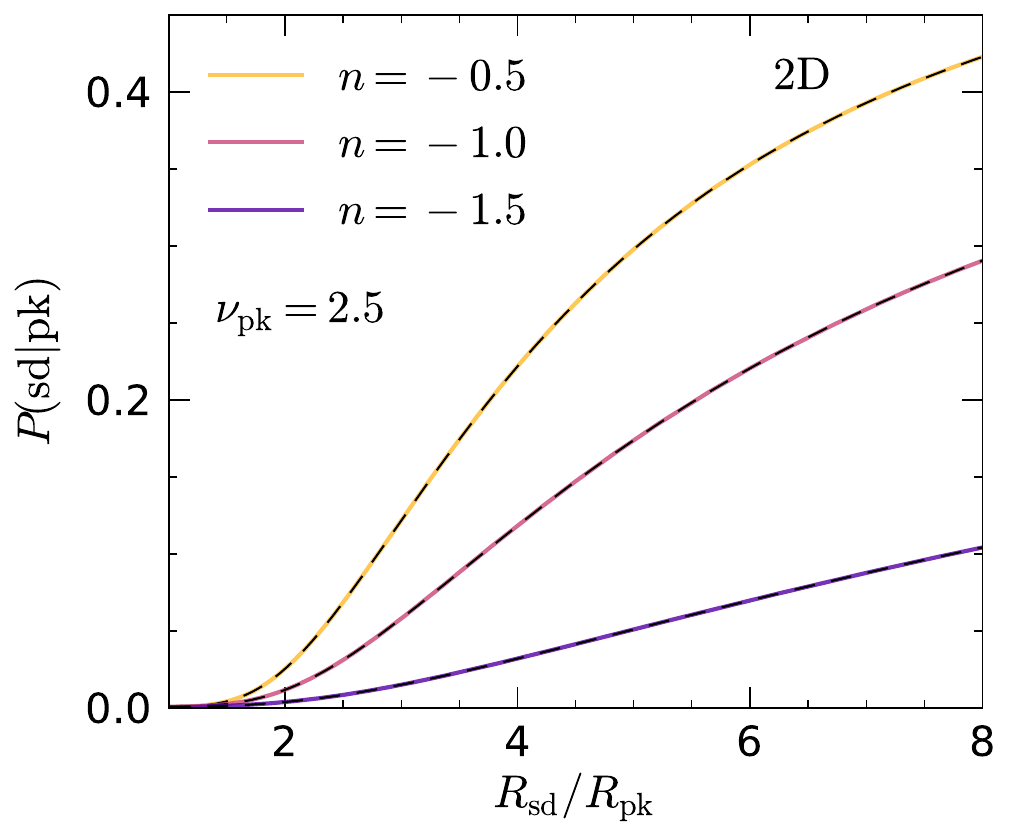}
\caption{Similar to Figure~\ref{fig:1}, but for $\nu_{\rm pk}=2.5$ in a two-dimensional GRF, computed using Equation~\eqref{eqn:condPDF2d}, with a power-law spectrum, $P_m(k)\propto k^{n}$. Solid curves show numerical results, whereas black dashed curves show analytic predictions. }\label{fig:4}
\end{figure}

We first show in Figure~\ref{fig:4} the conditional probability, $P(\mathrm{sd}|\mathrm{pk}) $, for power-law spectra, $P(k)\propto k^{n}$,
and at fixed $\nu_\mathrm{pk}=2.5$.
Our numerical calculations are in full agreement with the analytical predictions of Appendix~\ref{appen2}.
Similarly to Figure~\ref{fig:1}, we observe that the probability increases with the scale ratio, approaching a nearly constant value at very large $R_{\rm sd}/R_{\rm pk}$. For example, for $n=-0.5$, the conditional probability saturates at around $R_{\rm sd}/R_{\rm pk}\approx30$, earlier than  for the $n=-1$ and $-1.5$ cases. We also find a notable dependence on the spectral index of the power spectrum. At a fixed scale ratio, the probability becomes smaller for a power spectrum with more negative index, implying higher suppression of filamentary behaviour near the peaks\footnote{
Unlike the 3D case, where two distinct types of saddle points exist (including filament-type saddles), only a single class of saddles occurs in 2D. We nonetheless describe a region as {\it filamentary} if it collapses along one axis while expanding along the other. In two dimensions, such regions may be either over- or underdense. We therefore do not impose a positive overdensity constraint, which allows the 2D case to be treated on the same footing as the 3D one. By contrast, in 3D, filamentary regions are generally overdense even without such an explicit constraint.
}.  
This is due to the greater influence of long-wavelength modes, whose large-scale effects increase the likelihood that the surroundings of the peak will collapse simultaneously along all axes.

\begin{figure}[!t]
\centering
\includegraphics[clip,width=0.97\columnwidth]{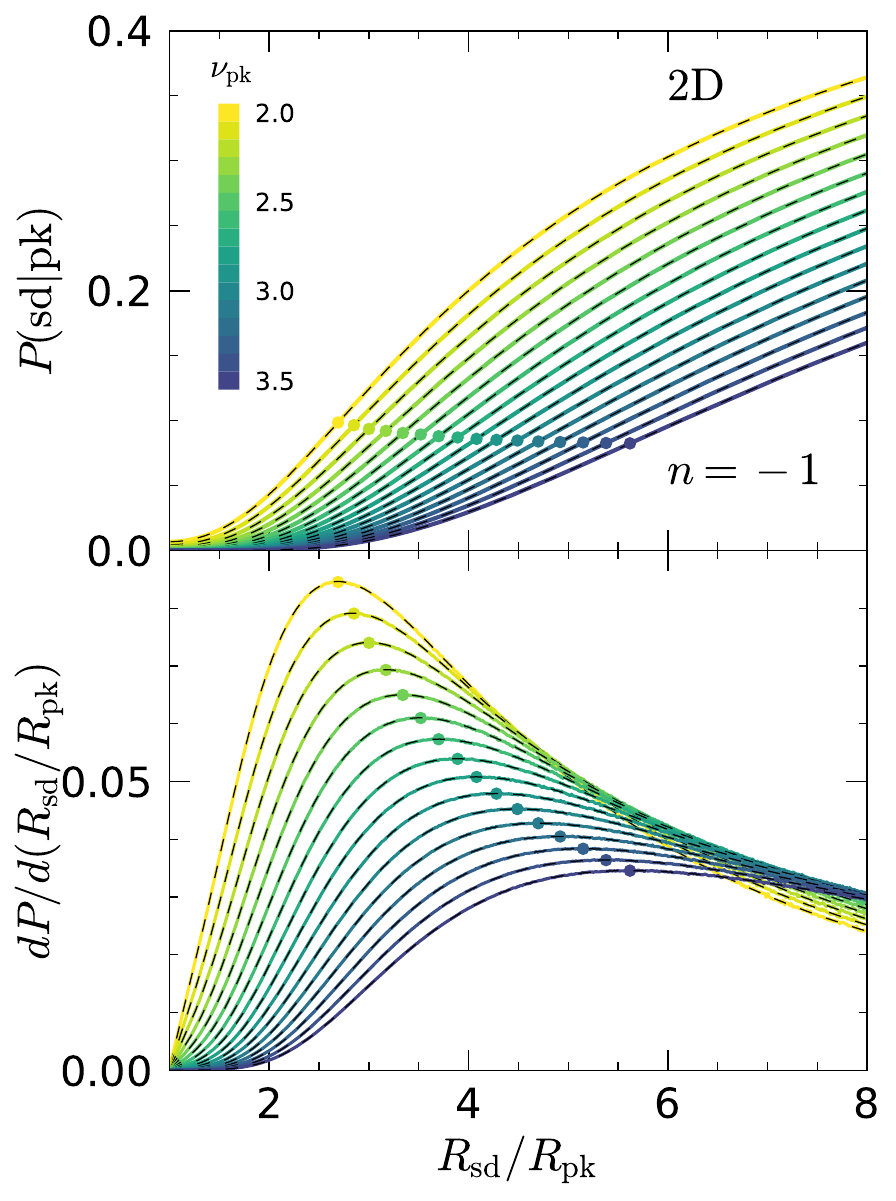}
\caption{Conditional probability (top) and its derivative (bottom) for a 2D GRF with power spectrum, $P_m(k)\propto k^{-1}$, computed using Equation~\eqref{eqn:condPDF2d}. Solid curves are numerical results, and black dashed curves show analytic predictions, for 15 equally-spaced $\nu_{\rm pk}$ as labeled. Inflection points (filled circles) are identified as maxima of the derivatives of  $P(\mathrm{sd}|\mathrm{pk}) $ with respect to the scale ratio.}\label{fig:5}
\end{figure}

In Figure~\ref{fig:5}, we display the dependence of the inflection point of the conditional PDF, $P(\mathrm{sd}|\mathrm{pk}) $, on peak rarity for the power spectrum with $n=-1$. Here we consider a wide rarity range, since the minimum rarity threshold for peak formation is slightly smaller in two dimensions than in the three-dimensional case.
The top panel indicates that, for a fixed peak size, higher density peaks are less likely to reside within the filamentary environments of the tidal field. This suggests that the conditions required for peak formation and two-dimensional compression become increasingly incompatible as peak rarity increases. Consequently, for more prominent peaks to achieve a comparable level of peak-saddle correlation, the tidal structures must be defined on significantly larger scales. In the bottom panel, inflection points are identified as maxima of the derivative. A clear trend emerges: the inflection point -- the location of highest sensitivity -- shifts toward larger scale ratios as the peak rarity increases.
\begin{figure}
\includegraphics[clip,width=\columnwidth]{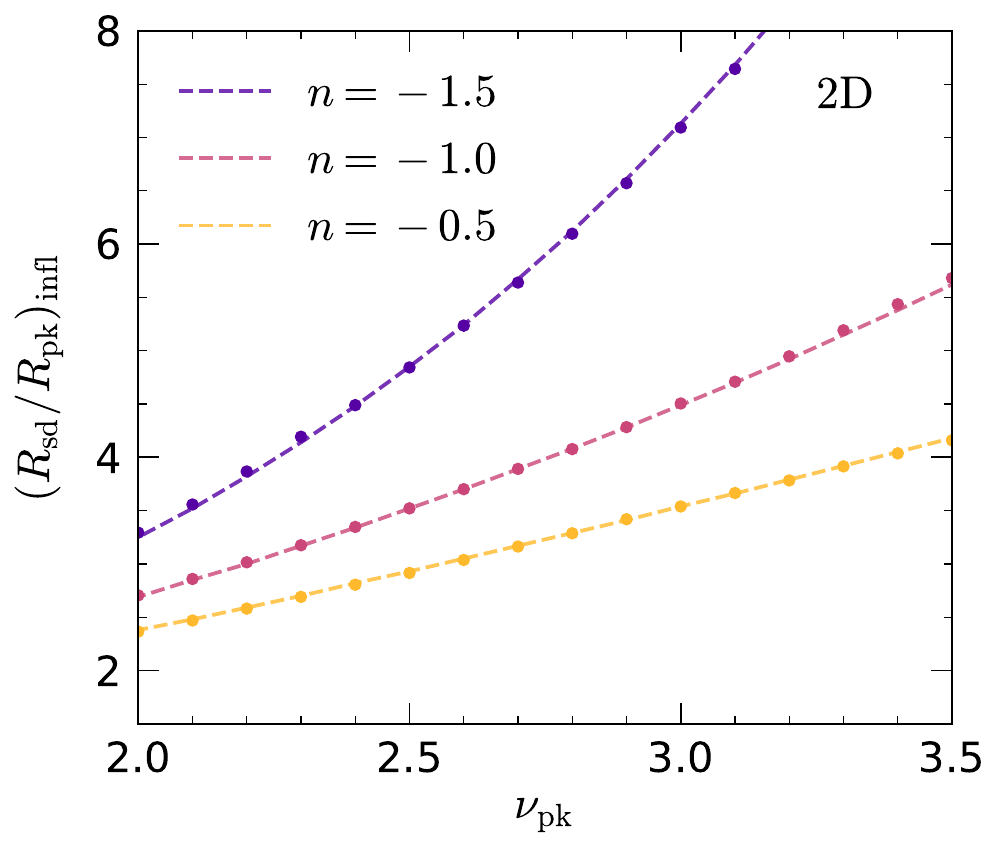}
\caption{Similar to Figure~\ref{fig:2}, but for peaks and saddles in a 2D GRF with power-law power spectra. Dotted lines represent numerical results, while dashed lines show analytical predictions. The peak-saddle correlation becomes most sensitive at larger scale ratios for rarer peaks.}\label{fig:6}
\end{figure}

Finally, Figure~\ref{fig:6} summarizes the increasing trend of filamentary scale with peak rarity in two dimensions. 
Our numerical results are in agreement with the analytical predictions. 
Interestingly, the relationship between the scale ratio and peak rarity becomes increasingly linear for higher spectral indices (e.g., $n \approx 0$), whereas it gradually transitions to a power-law behavior as the index approaches the negative of the spatial dimension (i.e., $n \sim -d$).

\subsection{3D Power-Law spectra}\label{appen4}
\begin{figure}
\includegraphics[clip,width=\columnwidth]{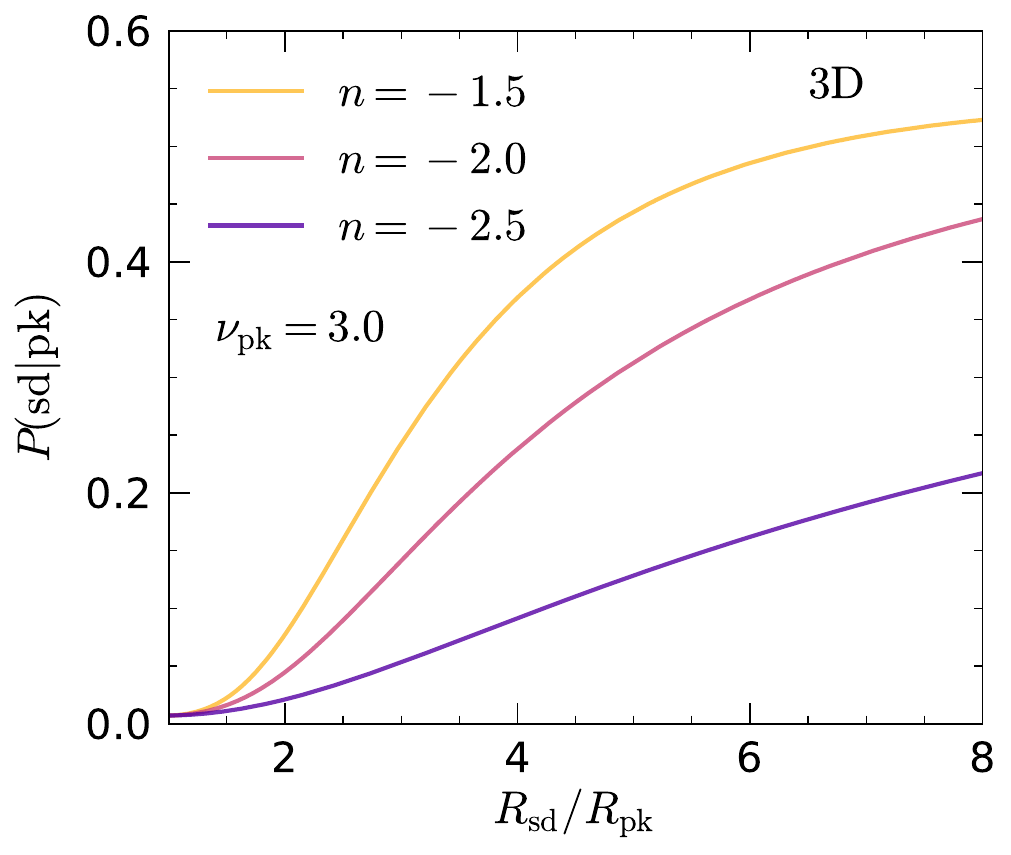}
\caption{Similar to Figure~\ref{fig:1}, but for peaks with rarity $\nu_{\rm pk}\ge3$ in a
3D GRF with power-law spectra, computed using Equation~\eqref{eq:condPDF}. Only numerical results are shown, because analytical expressions are not available in 3D.}\label{fig:7}
\end{figure}

Building on the consistency between the numerical and analytical results in two dimensions, we now extend our investigation to three dimensions, where the  derivation lacks a closed-form solution. Figure~\ref{fig:7} displays the numerically calculated conditional probabilities for power-law spectra in three dimensions. Overall, the three-dimensional results align  well with the two-dimensional trends: the halo-filament correlation strengthens with an increasing scale ratio, while it weakens as the spectral index becomes more negative.

\begin{figure}[!t]
\centering
\includegraphics[clip,width=0.97\columnwidth]{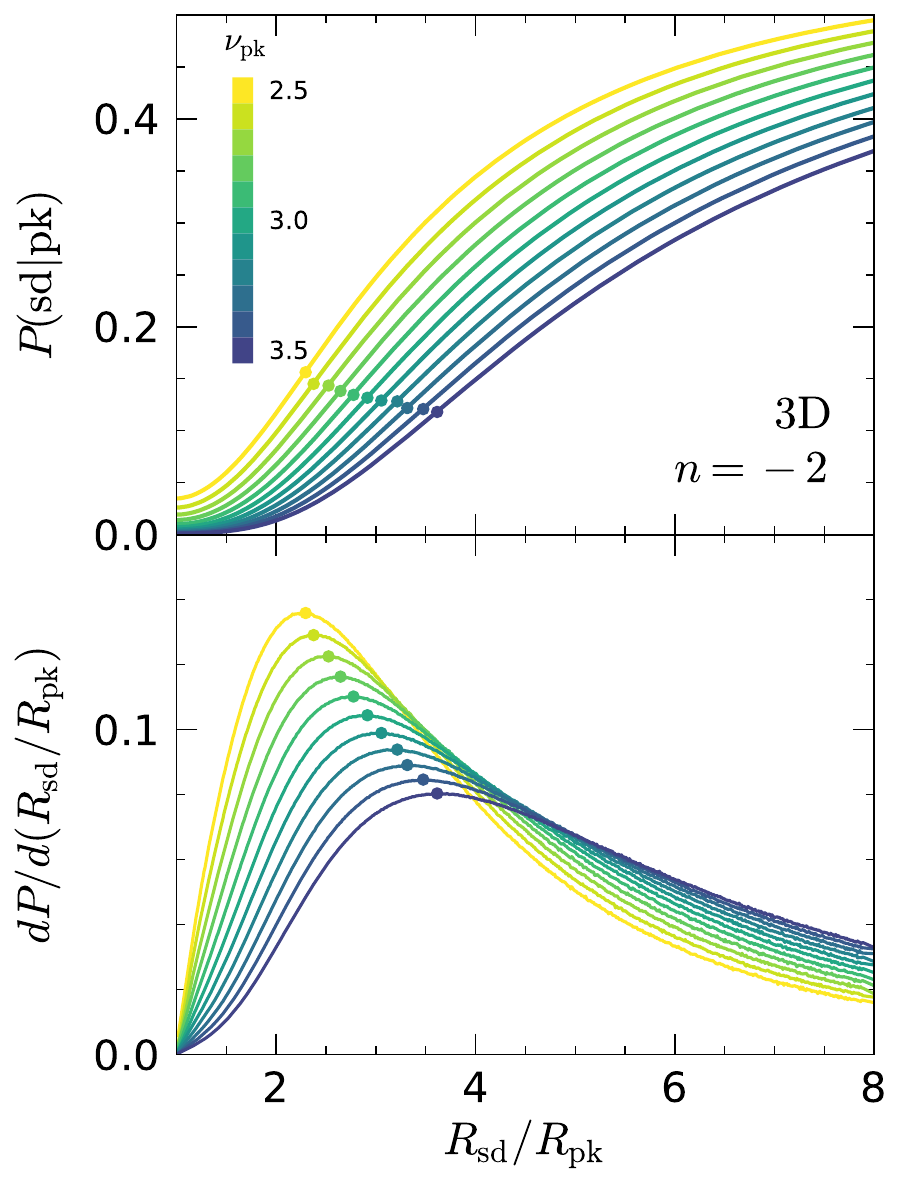}
\caption{Similar to Figure~\ref{fig:5}, but for a three dimensional GRF with a power law power spectrum, $P(k)\propto k^{-2}$. Only numerical results are shown, for 10 equally-spaced $\nu_{\rm pk}$. Qualitatively, the  realm of filaments for 3D power-law power spectras are consistent with those in 2D.}\label{fig:8}
\end{figure}
The peak rarity dependence of the conditional probability in three dimensions, shown in Figure~\ref{fig:8}, follows a trend analogous to the results for two dimensions presented in Figure~\ref{fig:5}. Specifically, for a fixed scale ratio, the halo-filament correlation weakens as peaks become more prominent. As shown in the lower panel, the characteristic scale ratio (defined by the apsis of the inflection point) shifts toward larger values for higher peaks. This trend reinforces the  earlier established scaling relation, confirming that more massive halos are most sensitively associated with an even larger filamentary environment.
%
\begin{figure}
\includegraphics[clip,width=\columnwidth]{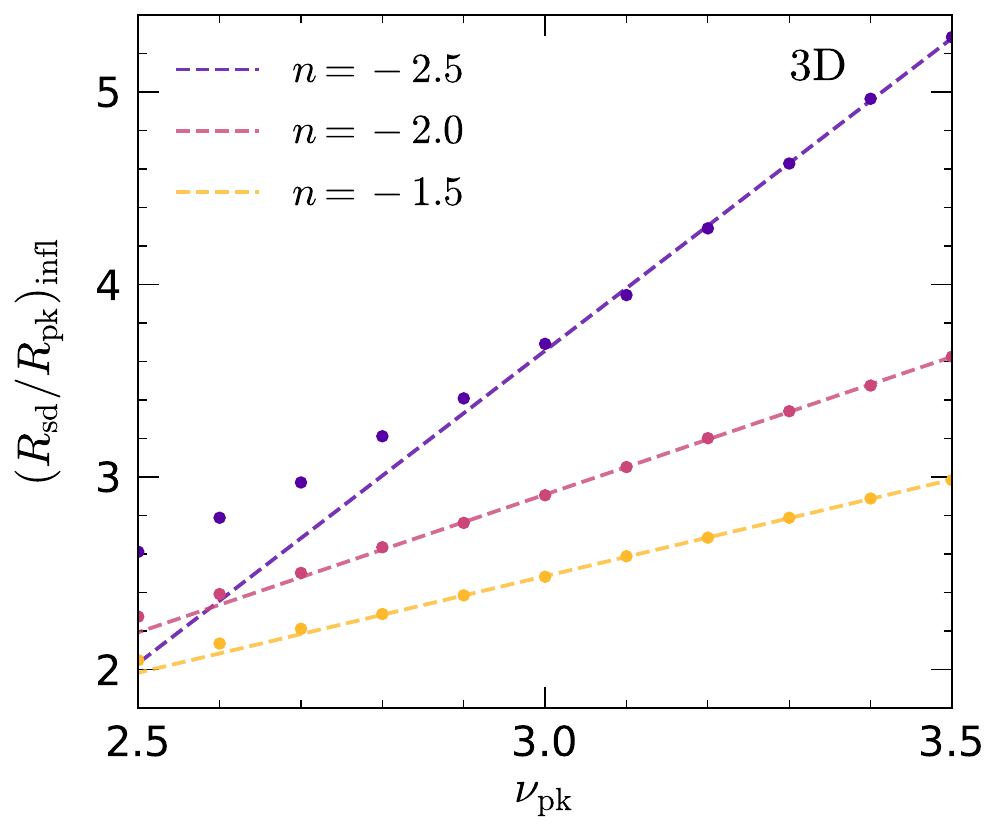}
\caption{Similar to Figure~\ref{fig:2}, but for power-law power spectra. The dots are numerical results, and dashed lines are linear fits derived from data points with $\nu_{\rm pk}\ge3$. Qualitatively, these power-law results for the realm of filaments are consistent with the $\Lambda$CDM power spectrum.}\label{fig:9}
\end{figure}
We again show the relation between the characteristic scale ratio and peak rarity in Figure~\ref{fig:9}. As for the three-dimensional $\Lambda$CDM power spectrum, we employ a linear fit calibrated for $\nu_{\rm pk}\ge3$, which provides a very good match to the numerical results. This linear fit remains valid even when extended to lower-rarities, except for $n=-2.5$, where we find a growing deviation from the fit with decreasing $\nu_{\rm pk}$.
These power-law results provide a theoretical benchmark for interpreting the $\Lambda$CDM results discussed in the main text (in Figures~\ref{fig:1} and ~\ref{fig:2}), and in the next Appendix. 

\section{  $\Lambda$CDM versus  Power-Law spectra}\label{app:lcdm}

The key distinguishing feature of the realistic 
$\Lambda$CDM model is the scale dependence of the effective spectral slope.
To facilitate comparison with power-law results, we can define an effective spectral index, $n_{\rm eff}$, which characterizes the slope of the $\Lambda$CDM power spectrum when smoothed on different scale $R$, 
\begin{equation}
    n_{\rm eff}\equiv -3 - \frac{\mathrm{d}{\rm log}\sigma^{2}(R)}{\mathrm{d}{\rm log}R}.
\end{equation}
Here, $\sigma^{2}(R)$ denotes the variance of matter density fluctuations smoothed with a Gaussian filter. For a power-law spectrum ${P_m(k) \propto k^{n}}$, this definition simply gives $n_{\mathrm{eff}} = n$.

\begin{figure}
\includegraphics[clip,width=\columnwidth]{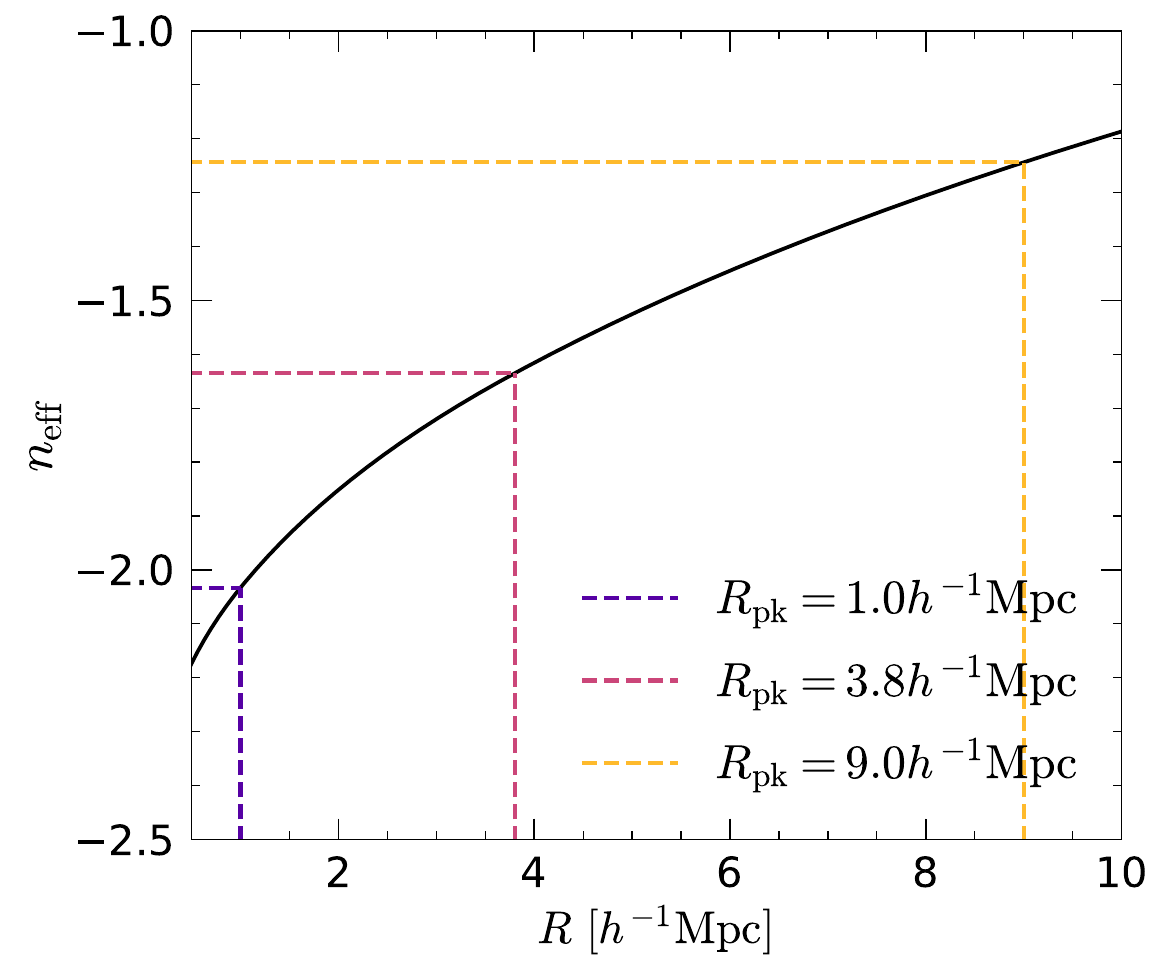}
\caption{Effective spectral index for a $\Lambda$CDM power spectrum as a function of the Gaussian smoothing scale. Dashed lines mark the effective spectral indices corresponding to the peak sizes we considered in this analysis. 
The middle value is equivalent to a $8h^{-1}{\rm Mpc}$ tophat smoothing.
The effective spectral index increases toward a larger smoothing scale.}\label{fig:neff}
\end{figure}

The relation between the Gaussian smoothing scale (in the units of $\mathrm{h^{-1} Mpc}$)  and the effective spectral index is shown in Figure~\ref{fig:neff},
and can be approximated\footnote{it is accurate over the range ${0.05 < R/h^{-1}(\mathrm{ Mpc) < 20}}$  of Gaussian scales.}
as
\begin{equation}
n_\mathrm{eff} \approx -2.35 + 0.36 \sqrt{R} + 0.064 \log R ~,
\label{eq:neff_fit}
\end{equation}
which provides a mapping between $\Lambda$CDM and power-law power spectra models. 
Since the $\Lambda$CDM power spectrum smoothed on a smaller scale yields more negative $n_{\rm eff}$, the observed trends -- higher conditional probabilities and smaller characteristic scale ratios for a larger peak -- are consistent with the behaviors shown in the power-law spectra analysis. We note that the enclosed masses of halos with top-hat radii corresponding to Gaussian $R_{\rm pk}=1.0, 3.8$, and $9h^{-1}{\rm Mpc}$ are $M=3.4\times10^{12}, 1.8\times10^{14}$ and $2.4\times10^{15}h^{-1}M_{\odot}$ at $z=0$, respectively.

\label{appen5}
Power law spectra also provide an analytic form for the redshift dependence of the filamentary scale around the objects collapsing
at a given $z$.  Given that on the one hand, $\sigma_0(R_\mathrm{pk})= \sigma_8 (R_8/R_\mathrm{pk})^{(n+d)/{2}}$, where $R_8$ is a normalization
scale, and, that on the other hand, $\sigma_0(R_\mathrm{pk}) = \delta_c/D(z) \nu_\mathrm{pk}$ from the criterion of collapse, we obtain
from Equation~\ref{eq:Rimplicit}
\begin{equation}
R_\mathrm{sd} =  R_8 \left( \frac{D(z) \sigma_8 \nu_\mathrm{pk}}{\delta_c \gamma_*}\right)^{\frac{2}{n+d}}
 \left(1+ \sqrt{1-\gamma_*^{\frac{4}{n+d}}} \right)\,.
\end{equation}
Thus, for the power-law spectra $R_\mathrm{sd} \propto D(z)^{{2}/(n+d)}$,  i.e.,  $R_\mathrm{sd} \propto (1+z)^{-{2}/(3+n)}$ in an Einstein-De Sitter universe.

\begin{figure}
\includegraphics[clip,width=\columnwidth]{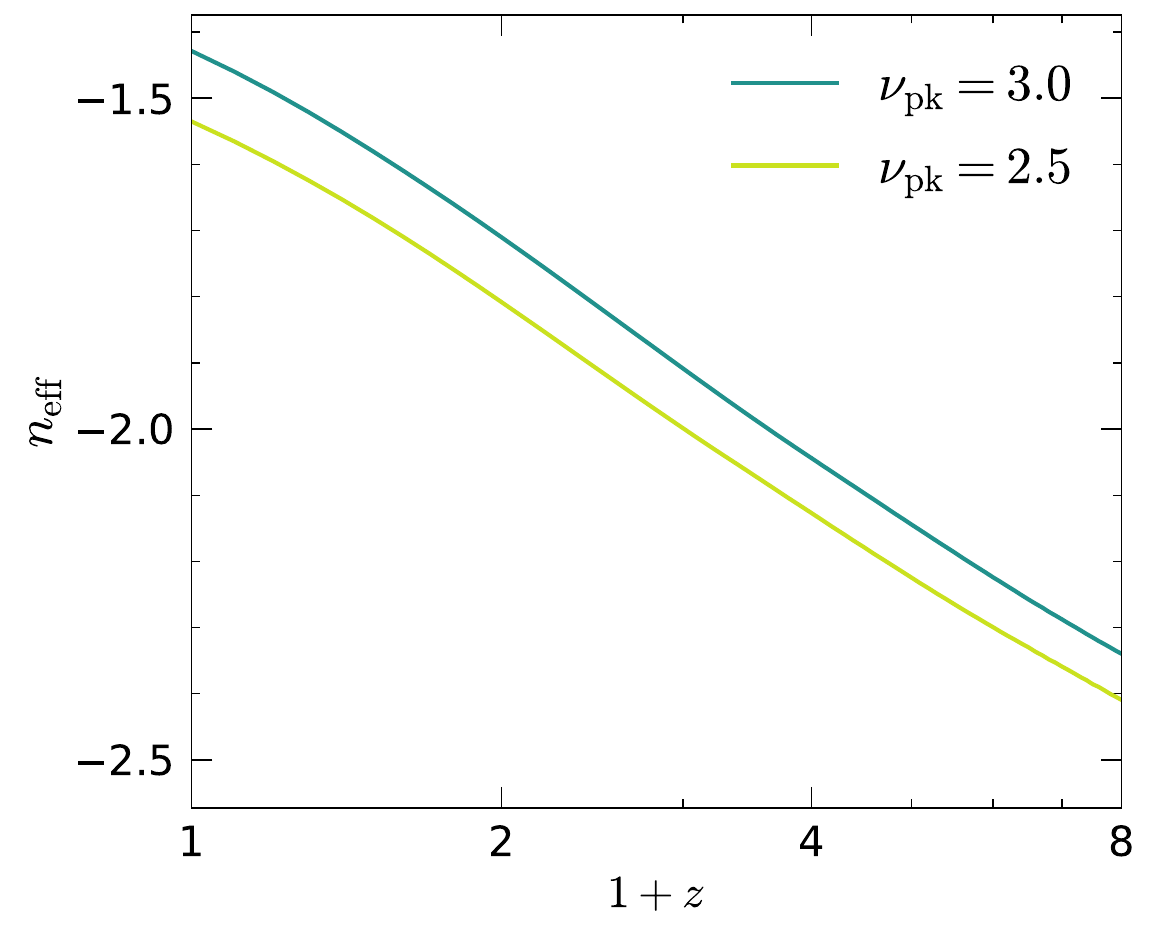}
\caption{Effective spectral index for a $\Lambda$CDM power spectrum at the scale of the peaks with $\nu_{\rm pk}=3.0$ and $2.5$, collapsed at redshift $z$. The effective spectral index decreases toward higher redshifts.}\label{fig:11}
\end{figure}

In a $\Lambda$CDM universe, the spectrum is not scale-invariant, as illustrated in Figure~\ref{fig:11}, and the growing mode likewise departs from $(1+z)^{-1}$.
Therefore, here we provide a fitting form for the characteristic scale of filaments as a function of redshift in $\Lambda$CDM cosmology, which remains accurate to within $3\%$:
\begin{equation}
\hskip -0.25cm R_{\rm sd} [h^{-1} \mathrm{Mpc}]= 
\begin{cases} 
\displaystyle \frac{43-3.3(1+z)}{2.8+(1+z)^2}\!\! & \text{for } \nu_{\rm pk}=2.5,\\[0.5ex]
\displaystyle \frac{76.4-4.6(1+z)}{3.7+(1+z)^2}\!\! & \text{for } \nu_{\rm pk}=3.0, 
\end{cases} \label{eq:Rdzprecise}
\end{equation}

In closing, the power-law results presented in Appendices~\ref{appen2}, \ref{appen3} and~\ref{app:lcdm} robustly validate our main finding presented in Equations~\eqref{eqn:linear}, \eqref{eq:z0fit}, \eqref{eq:massfil} and \eqref{eq:Rdcase}.

\end{document}